\documentstyle[preprint,aps]{revtex}
\begin{document}
\draft

\title
{On a class of integrable interacting electronic systems with off-diagonal
disorder}
\author{A.~H.~Castro Neto$^{1,2}$, D.K. Campbell$^{2}$,
and Philip Phillips$^{2}$}
\bigskip
%
\address
{$^{1}$ Institute for Theoretical Physics\\
University of California\\
Santa Barbara, CA, 93106-4030}

\address
{$^{2}$ Loomis Laboratory of Physics\\
University of Illinois at Urbana-Champaign\\
1100 W.Green St., Urbana, IL, 61801-3080}
%
\maketitle

\begin{abstract}
We report a class of {\it integrable} one-dimensional interacting electronic
systems
with {\it off-diagonal disorder}. For these systems,
the disorder can be ``gauged away,''
and the spectrum can be mapped completely onto
the spectrum of the ordered problem, which can be solved by Bethe ansatz
or by bosonization. We study the magnetic properties of the persistent
currents in mesoscopic rings in the case in which the ordered system
is a Luttinger liquid. The system can be paramagnetic
or diamagnetic, depending on the amount of the disorder and the
number of fermions in the system.
\end{abstract}

\bigskip


\narrowtext

It is usually believed that any amount of disorder in one-dimensional
systems leads to electron localization. This belief is based
mainly on the work of Anderson \cite{anderson} and Mott \cite{mott},
which demonstrates that in a one dimensional,
non-interacting electronic system, disorder leads to destructive interference
and as a consequence localization of the electronic wave functions.
Recently, a class of 1-dimensional models has been discovered in which at
certain
energies, localization is suppressed as a result of a subtle resonant effect.
The random dimer model \cite{phillip}, used for instance
to explain the high conductivity of doped polyaniline, \cite{phillip2},
is the archtypeal model here.
When interactions are included in the presence of
disorder, even more complex behaviour can arise.
In some cases, such as
$\nu=2/3$ edge states in the fractional quantum Hall effect, the disorder
appears to
act as an irrelevant operator \cite{kfp}. In addition, it has
been shown at the level of mean-field theory that interactions might destroy
resonant scattering in 1-dimensional systems \cite{ortiz}. This issue remains
controversial as only a mean-field answer has been obtained.

In this paper we present an {\it integrable, interacting model with disorder}.
We consider a particle hopping on a one-dimensional chain with
a hopping matrix element that has a random phase. That
the matrix element is complex means that time-reversal symmetry
has been broken. Such a situation can arise, for instance, in
the presence of strong spin orbit coupling \cite{soc}. A similar type
of problem was studied in the late seventies by Mattis \cite{mattis}
and by Luttinger \cite{luttinger} in the context of the Ising model.
Apart from its intrinsic interest as an integrable system, our model
is relevant to understanding the problem
of persistent currents in mesoscopic rings. As we will show,
the zero temperature magnetic properties of the disordered
integrable system are quite remarkable: the system can be paramagnetic or
diamagnetic,
depending on both the number of fermions (an effect that has been
studied previously \cite{loss}) and the amount of disorder in the system.

We will consider the problem of interacting electrons with charge $e$
moving in a ring (periodic boundary conditions) with $N_a$ sites
which is subject to an external magnetic flux $\Phi$.
The Hamiltonian of interest is
\begin{equation}
H = - \sum_{i=1,\sigma}^{N_a} t_{i,i+1} e^{i \frac{2 \pi \Phi}{N_a \Phi_0}}
\left( c^{\dag}_{i,\sigma}
c_{i+1,\sigma} + c^{\dag}_{i+1,\sigma} c_{i,\sigma}\right) +
{\cal V} \lbrack n_{i,\sigma} \rbrack
\end{equation}
where $c_{i,\sigma}$ and  $c^{\dag}_{i,\sigma}$
are the annihilation and
creation operators for an electron in the $i^{th}$ site with spin projection
$\sigma$ ($\sigma = \uparrow$ or $\downarrow$), ${\cal V} \lbrack n_{i,\sigma}
\rbrack$
describes the interaction between the electrons (which is assumed to
depend only on the electron {\it density}, {\it i.e.},
number of electrons on a site ($n_{i,\sigma} =
c^{\dag}_{i,\sigma} c_{i,\sigma}$)), $t_{i,i+1}$ is the hopping matrix element
(which will be the source of the disorder), and $\Phi_0 = \frac{h}{e}$ is the
flux quantum. This Hamiltonian is quite general in terms of the
electron-electron
interaction; for instance, it can describe Hubbard models involving
only an on-site interaction,
${\cal V} \equiv {\cal V}_1 = U \sum_i n_{i,\uparrow} n_{i,\downarrow}$,
extended Hubbard models having also nearest-neighbor interactions,
$ {\cal V} = {\cal V}_1 + V \sum_{i,\sigma} n_{i,\sigma} n_{i+1,\sigma}$,
{\it etc.}.
The disorder in the  model arises from the hopping matrix element,
which is chosen to be,
\begin{equation}
t_{i,i+1} = t e^{i \delta_i}
\end{equation}
where $t$ is a constant and $\delta_i$ is a random phase on the site which
can vary between $-\pi < \delta_i \leq \pi$.

Becasue the disorder appears as just a phase in the Hamiltonian, it can
be (almost !) completely ``gauged away'' by an unitary transformation,
\begin{equation}
{\cal U} = \exp\left\{-i \sum_{j=1}^{N_a}  \sum_{n=1,\sigma}^{N_a}
\sum_l^{n-1} \left(\delta_l+\frac{2 \pi \Phi}{N_a \Phi_0}\right) \, \,
c^{\dag}_{j,\sigma} c_{j,\sigma} \right\}
\end{equation}
which transforms the fermion operator to
\begin{equation}
d_{j,\sigma} =
{\cal U} c_{j,\sigma} {\cal U}^{\dag} = \exp\left\{i \left(\sum_{l=1}^{j-1}
\delta_l
+(j-1) \frac{2 \pi \Phi}{N_a \Phi_0}\right)\right\} c_{j,\sigma}.
\end{equation}
Under this unitary transformation it is easy to see that the new
Hamiltonian can be written as,
\begin{equation}
H'= {\cal U} H {\cal U}^{\dag}  = - t \sum_{i=1,\sigma}^{N_a} \left(
d^{\dag}_{i,\sigma}
d_{i+1,\sigma} + d^{\dag}_{i+1,\sigma} d_{i,\sigma}\right) +
{\cal V} \lbrack n_{i,\sigma} \rbrack,
\end{equation}
since $ {\cal U} n_{j,\sigma} {\cal U}^{\dag} =  n_{j,\sigma}$. Eqn. (5)
describes the underlying {\it ordered} problem, simply reexpressed in
terms of the $ d_{j,\sigma}$; thus we conclude that (modulo boundary
conditions, as we discuss below)
the integrability of the disordered problem depends only on the
integrability of the ordered problem (5). Here we will consider cases
in which Eqn. (5) can be solved by the Bethe ansatz \cite{bethe} or by
bosonization
\cite{bosonization}.

Because we are treating a ring,
we choose periodic boundary conditions,
that is, $ c_{j,\sigma} = c_{j+N_a,\sigma}$. In terms of the new
operators, this requires
\begin{equation}
d_{j+N_a,\sigma} = \exp\left\{i\left(\Delta+2 \pi
\frac{\Phi}{\Phi_0}\right)\right\} d_{j,\sigma}
\end{equation}
where $\Delta = \sum_{l=1}^{N_a} \delta_l$. Note that $\Delta$ can
be always written as
\begin{equation}
\Delta = 2 \pi n + \delta
\end{equation}
where $n$ is an integer and $-\pi < \delta \leq \pi$. Observe that
the effect of the disorder is to create an {\it effective flux},
$\Phi_{eff}$, through the ring. This flux is given by
\begin{equation}
\Phi_{eff} = \Phi + n \Phi_0 + \frac{\Phi_0 \delta}{2 \pi}.
\end{equation}
Thus, disorder in this problem is effectively equivalent
to an increase in the magnetic field applied in the ring. Without
loss of generality, we will assume that $n=0$ in what follows.

It is interesting to look at the wavefunction of the
original disordered problem, $|\Psi>$, in terms of the wavefunction of
the underlying ordered problem, $|\Psi'>$, which is an eigenstate of the
Hamiltonian (5). Under the unitary transformation (3) we have,
\begin{equation}
|\Psi> = U^{\dag} |\Psi'>.
\end{equation}
The general form for the wavefunction for the ordered system is
\begin{equation}
|\Psi'> = \prod_{l=1}^{N} \sum_{j_l,\sigma_l} \psi_{\sigma_1,
\sigma_2,\cdots,\sigma_N}(j_1,j_2,\cdots,j_N) \, \, d^{\dag}_{j_1,\sigma_1}
d^{\dag}_{j_2,\sigma_2} \cdots d^{\dag}_{j_N,\sigma_N}|0>
\end{equation}
where $N$ is the number of electrons in the lattice, $|0>$
is the vacuum state of the problem.

Let us consider first the case where (5) can be solved by the Bethe ansatz
with the twisted boundary conditions (6). In this case the wave amplitude
$\psi$ can be written as \cite{bethe}
\begin{equation}
\psi_{\sigma_1,\sigma_2,\cdots,\sigma_N}(j_1,j_2,\cdots,j_N) = \sum_{{\cal P}}
A_{\sigma_1,\sigma_2,\cdots,\sigma_N}({\cal Q}|{\cal P}) e^{i \sum_{l=1}^{N}
k_{{\cal P}_l} j_l}
\end{equation}
for a specific ordering
$\{1<j_{{\cal Q}1}<j_{{\cal Q}2}<...<j_{{\cal Q}N}<N_a\}$ of
the particles. $\cal{P}$ and $\cal{Q}$ are permutations of the coordinates
$j_l$ and momenta $k_l$ of the electrons (we measure momentum in units of
the inverse lattice constant) and
\begin{equation}
A_{\sigma_1,\sigma_2,\cdots,\sigma_N}({\cal Q}|{\cal P}) =
(-1)^{\eta_{{\cal P}}} A_{\sigma_{{\cal Q}_1},\sigma_{{\cal Q}_2},
\cdots,\sigma_{{\cal Q}_N}}({\cal Q}| {\cal P})
\end{equation}

are the amplitudes obtained by the Bethe ansatz equations for the ordered
case (the factor $\eta_{{\cal P}}$ is $0$ or $1$ if the permutation is
even or odd, respectively).

The effect of disorder in the wave function
can be obtained from (9) by introducing the
identity operator between the operators in (10) ($U U^{\dag} =1$).
It is straightforward to show that the wavefunction of the disordered
system is simply written as ($\Phi =0$),
\begin{equation}
|\Psi> = \prod_{l=1}^{N} \sum_{j_l,\sigma_l} \tilde{\psi}
_{\sigma_1,\sigma_2,\cdots,\sigma_N}(j_1,j_2,\cdots,j_N)
c^{\dag}_{j_1,\sigma_1}
c^{\dag}_{j_2,\sigma_2} \cdots c^{\dag}_{j_N,\sigma_N}|0>
\end{equation}
where
\begin{equation}
\tilde{\psi}_{\sigma_1,\sigma_2,\cdots,\sigma_N}(j_1,j_2,\cdots,j_N)=
\psi_{\sigma_1,\sigma_2,\cdots,\sigma_N}(j_1,j_2,\cdots,j_N)
\exp\left(i \pi \sum_{l=1}^N N_l \delta_l\right)
\end{equation}
where $N_l = N-m$ for $j_{{\cal Q}m} \leq l \leq j_{{\cal Q}(m+1)}-1$
in the ordering defined above. Thus, we conclude that the effect
of this particular kind of disorder is to introduce a Berry phase
in the wavefunction \cite{berry}. Therefore, the disorder will introduce
fluctuations in the shape of the wavefunction.
For example, for the non-interacting
problem, instead of an extended Bloch wave, we would find an extended
but strongly oscillating solution. Of course the solution of the problem
now depends on
the {\it specific} form of the Hamiltonian which is solved by the Bethe
ansatz with the twisted boundary conditions.

As an important example of the relevance this model, let us consider
the transport properties of this system at low energies.
Using a well-known result due
to Kohn \cite{kohn}, we write the charge stiffness of this system as
\begin{equation}
D_c = \frac{N_a}{2} \left(\frac{d^2 E_0(\Phi)}{d \Phi^2}\right)_{\Phi=0},
\end{equation}
where $E_0(\Phi)$ is the ground state energy of the system in
the presence of the magnetic flux. $E_0(\Phi)$ can
be calculated from the Bethe ansatz solution \cite{flux,millis}.
To explore the implications for transport in detail, let us take
the continuum limit of the Hamiltonian (5), which (for typical
band fillings) is a Luttinger model. For simplicity, let us consider the
specific case of spinless fermions which interact via
\begin{equation}
{\cal V} \lbrack n_{i} \rbrack = U \sum_{i=1}^{N_a}
\left(n_i-\frac{1}{2}\right)
\left(n_{i+1}-\frac{1}{2}\right).
\end{equation}

This problem has been solved for an {\it external} flux \cite{loss},
using the bosonization technique with twisted boundary conditions.
We can apply the results of \cite{loss} directly, substituting
for the external flux the {\it effective} flux due to
disorder, as defined in (8). With some trivial modifications of the
formula for $D_c$ in ref. \cite{loss} one finds, at zero temperature
and for an odd number of electrons,
\begin{eqnarray}
D_c &=& D_c^0 \, \sum_{n=1}^{\infty} (-1)^{n+1} \frac{\cos(n \delta)}{n}
\nonumber
\\
    &=& D_c^0 \, \ln \left(2 \cos \frac{\delta}{2}\right)
\end{eqnarray}
and for an even number of electrons,
\begin{eqnarray}
D_c &=& - D_c^0 \sum_{n=1}^{\infty} \frac{\cos(n \delta)}{n}
\nonumber
\\
    &=& D_c^0 \ln \left(2 (1-\cos \delta)\right)
\end{eqnarray}
where $D_c^0 = \frac{v_F^*}{\pi}$ and $v_F^*$ is the velocity of
propagation of the fermions renormalized by the interactions \cite{loss}.
For an odd number of electrons,
the system is diamagnetic if $|\delta| < \frac{2 \pi}{3}$
and is paramagnetic if $\frac{2 \pi}{3} < |\delta| \leq \pi$.
For an even number of electrons, the system is diamagnetic
for $\frac{\pi}{3} < |\delta| \leq \pi$ and paramagnetic for
$|\delta| < \frac{\pi}{3}$. Thus, the magnetic
response of the system is very sensitive to both the ``parity'' ($N$) and
the amount of disorder ($\delta$) in the system.

The above argument is valid for a fixed value of $\delta$. Suppose,
however, that we have an ensemble of rings with different values
of the disorder. We asume now that $\delta$ is randomly distributed between
$-\alpha$ and $\alpha$ where $\alpha \leq \pi$, that is, the probability of
finding an amount of disorder between $\delta$ and $\delta + d \delta$
is given by
\begin{eqnarray}
P(\delta) d \delta = \frac{1}{2 \alpha} \Theta(\alpha - |\delta|) d \delta
\end{eqnarray}
where $\Theta$ is the usual Heavyside step function.
In this case, from
(8) we would find that the average effective flux in the ring equals
the applied flux, that is, $\bar{\Phi}_{eff} = \Phi$.
It is easy to
show by averaging (17) and (18) with (19) that, for an odd number of
electrons one finds,
\begin{equation}
\bar{D}_c = D_c^0 \frac{1}{\alpha} \sum_{n=1}^{\infty} \frac{(-1)^{n+1}}{n^2}
\sin n \alpha
\end{equation}
and the response is always diamagnetic and for an even number
of electrons,
\begin{equation}
\bar{D}_c = - D_c^0 \frac{1}{\alpha} \sum_{n=1}^{\infty} \frac{1}{n^2}
\sin n \alpha
\end{equation}
and the response is always paramagnetic. This behavior
resembles the problem without disorder. However, it is easy to
see that for $\alpha = \pi$ the charge stiffness vanishes in
both cases. This is due to the random distribution of diamagnetic
and paramagnetic rings which gives rise to an effect that averages
to zero. This is clear in (17) and (18) since the stiffness is
periodic in $\delta$.

In summary, we have presented in this paper a class of integrable
disordered one-dimensional (periodic) systems in which
the disorder can be ``gauged away''.
In such systems, the integrability only depends on the integrability
of the underlying ordered problem. For problems solvable by the
Bethe ansatz, the wavefunction acquires a Berry phase due to the motion
of the electrons in the disordered system.
We show that the disorder acts
as an effective flux through the ring and that, for a fixed total amount
of disorder, beyond the usual
sensitivity of the persistent current to the number of electrons in
the system, the magnetic response depends also on the amount of disorder in
the system.

We are deeply indebt to W.E.Goff for many discussions.
We thank N. Argaman, M.P.A. Fisher and E. Fradkin for
clarifying discussions,
and D. Guo for sharing her computer simulations of a
modified Hubbard model which led us to the recognition
that this class of models might be integrable. We
acknowledge the partial support of by NSF Grants DMR94-96134,
DMR91-22385 and PHY89-04035.


\begin{references}

\bibitem{anderson}
P. W. Anderson, {\it Phys. Rev.} {\bf 109}, 1492, (1958).
\bibitem{mott}
N. F. Mott and W. D. Twose, {\it Adv. Phys.} {\bf 10}, 107, (1961).
\bibitem{phillip}
D. H. Dunlap, H. L. Wu and P. Phillips, {\it Phys. Rev. Lett.} {\bf 65}, 88,
(1990).
\bibitem{phillip2}
H. L. Wu and P. Phillips, {\it Phys. Rev. Lett.} {\bf 66}, 1366, (1991);
P. Phillips and H. L. Wu, {\it Science} {\bf 252}, 1805, (1991).
\bibitem{kfp}
C. L. Kane, M. P. A. Fisher and J. Polchinski, {\it Phys. Rev. Lett.}
{\bf 72}, 4129, (1994).
\bibitem{ortiz}
P. Ordejon, G. Ortiz and P. Phillips, {\it Phys. Rev. B} {\bf 68}, 14682,
(1994).
\bibitem{soc}
S. Fujimoto and N. Kawakami, {\it Phys. Rev. B} {\bf 48}, 17406, (1993) and
references therein.
\bibitem{mattis}
D. C. Mattis, {\it Phys. Lett.} {\bf 56 A}, 421, (1976).
\bibitem{luttinger}
J. M. Luttinger, {\it Phys. Rev. Lett.} {\bf 37}, 778, (1976).
\bibitem{loss}
D. Loss, {\it Phys. Rev. Lett.} {\bf 69}, 343, (1992).
\bibitem{bethe}
See, for instance, J. H. Lowenstein in {\it Recent Advances in Field
Theory and Statistical Mechanics - Les Houches, XXXIX, 1982}, J. B. Zuber
and R. Stora, eds. Elsevier, (1984).
\bibitem{bosonization}
V. J. Emery pp. 247-303 in {\it Highly Conducting One Dimensional Solids},
J. T. Devreese, R. P. Evrard and V. E. Doren, eds. Plenum, (1979).
\bibitem{berry}
M. V. Berry, {\it Proc. Roy. Soc. Lond. A}, {\bf 392}, 45, (1984).
\bibitem{kohn}
W. Kohn, {\it Phys. Rev.} {\bf 133}, A171, (1964).
\bibitem{flux}
B. S. Shastry and B. Sutherland, {\it Phys. Rev. Lett.} {\bf 65}, 243, (1990).
\bibitem{millis}
C. A. Stafford and A. J. Millis, {\it Phys. Rev. B} {\bf 48}, 1409, (1993).

\end{references}
\end{document}